\documentclass{aa}  

\usepackage{graphicx}
\usepackage{txfonts}
\usepackage{ulem}
\begin{document}

   \title{The extension of the Fundamental Metallicity Relation beyond the BPT star-forming sequence: evidence for both gas accretion and starvation}


\titlerunning{FMR extension to BPT-non-SF galaxies}

	\author{Nimisha Kumari
	\inst{1,2,3}\fnmsep\thanks{E-mail: kumari@stsci.edu (NK)},
	Roberto Maiolino\inst{2,3},
	James Trussler\inst{2,3},
	Filippo Mannucci\inst{4},
		Giovanni Cresci\inst{4},
		Mirko Curti\inst{2,3}
		Alessandro Marconi\inst{4}
			\and
			Francesco Belfiore\inst{4}
}

\institute{$^1$ AURA for the European Space Agency, Space Telescope Science Institute, 3700 San Martin Drive, Baltimore, MD 21218, US, \\
	$^2$ Kavli Institute for Cosmology, University of Cambridge CB3 0HA, UK, \\
	$^3$ Cavendish Laboratory, University of Cambridge CB3 0HE, UK, \\
	$^4$ INAF - Osservatorio Astrofisico di Arcetri, Largo E. Fermi 5, I-50125, Firenze, Italy\\
}

\authorrunning{Kumari, Maiolino, Trussler et al.}

   \date{Received ; accepted }

 
  \abstract
   {The fundamental metallicity relation (FMR) of galaxies is a 3D relation between the gas-phase metallicity, stellar mass  and star-formation rate (SFR). It has been studied so far only for galaxies identified as star-forming (SF) on the BPT diagrams (BPT-SF), but not for  galaxies with LI(N)ER/AGN classification (BPT-non-SF), mainly due to the lack of diagnostics for estimating their gas-phase metallicities in the latter cases.}
   {We extend the FMR to BPT-non-SF galaxies.}
   {To this end, we exploit the recent nebular line empirical calibrations derived specifically for galaxies classified as non-SF in the BPT diagrams. Moreover, we study an alternative representation of the FMR where we consider the offsets in metallicity and SFR with respect to Main Sequence (MS) galaxies.}
   {We find that galaxies with SFR higher than the MS are more metal-poor than their counterparts on the MS, which is interpreted in terms of gas accretion, boosting star formation and diluting the metallicity. Low-mass galaxies below  the MS (i.e. towards quiescence) have metallicities higher than their MS counterparts, which is interpreted in terms of starvation,  (i.e. suppression of fresh gas supply) hampering star formation and reducing the dilution effect, hence resulting in a higher level of internal chemical enrichment. Massive galaxies below the MS  have gas  metallicity much closer to their MS counterparts and much lower than expected from  their stellar metallicities; this result  suggests  a scenario where  massive nearly-quiescent galaxies with  LI(N)ER-like nebular emission have recently accreted gas from the circum/intergalactic medium.}
   {}


   \maketitle
%

 \section{Introduction}
 \label{section:introduction}
 
 \indent 
 The metallicity scaling relations of galaxies provide important information on the galaxy evolutionary processes and tight constraints to models and numerical cosmological simulations 
 \citep{Maiolino2019}. The tightest scaling relation is between the stellar mass of galaxies and their gas-phase metallicity, the so-called mass-metallicity relation (\citealp[MZR,][]{Lequeux1979}; \citealp[also see][]{Tremonti2004}), which is often interpreted in terms of the deeper gravitational potential well of massive galaxies being more effective in retaining metals than low-mass galaxies. However, additional secondary relation of the gas metallicity with other galactic properties have been found.
 \citet{Mannucci2010} utilised the vast dataset of star-forming (SF) emission line galaxies from the seventh data release \citep{Abazajian2009} of the Sloan Digital Sky Survey \citep[SDSS,][]{York2000} to explore the relationship between the galactic stellar mass (M$_{\star}$), star-formation rate (SFR) and gas-phase metallicity (12+log(O/H)). Following the works of \citet{Ellison2008}, they identified a secondary relation according to which, at a given stellar mass, the gas-phase metallicity of galaxies anticorrelated with the SFR. Such anticorrelation has been often interpreted in terms of gas accretion from the inter-galactic/circum-galactic medium which dilutes the gas metallicity and foster star formation. \citet{Mannucci2010} showed that the resulting three dimensional (3D) relation between stellar mass, SFR and metallicity 
 exhibited little scatter in metallicity.
 This three-dimensional relation was also found not to evolve in redshift, and was  dubbed the fundamental metallicity relation (FMR).

 \indent Several studies based on observations \citep[e.g.][]{Mannucci2011, Cresci2012, BrisbinHarwit2012, Yates2012, Roseboom2012, AndrewsMartini2013, NakajimaOuchi2014, Salim2014, Lian2015, Hunt2016, Kashino2016, Telford2016, Cresci2019, Curti2020}, simulations \citep[e.g.][]{Dave2011, DeRossi2015, LagosC2016, Genel2016, DeRossi2017, MattheeSchaye2018, Torrey2018, Bustamente2018} and analytical models \citep[e.g.][]{Dayal2013, Lilly2013, Forbes2014, Spitoni2017} have explored various aspects of the FMR. Though most observational studies agree on the general shape of the FMR, there have been discrepancies on the FMR behaviour typically at high stellar masses which are postulated to depend on various systematic effects such as selection of galaxies on the basis of signal-to-noise (S/N) ratio in emission lines \citep[e.g.][]{Yates2012, Telford2016} and calibration methods of SFR and gas-phase metallicities \citep[e.g.][]{AndrewsMartini2013, Cresci2019}. The overall normalization of the FMR is shown to depend critically on adopted metallicity calibration \citep{Curti2020}. The physics of the FMR is well explained by analytical models \citep[e.g.][]{Dayal2013, Lilly2013} whereby different components of the gas-regulator model play significant roles in shaping the form of the FMR.
 
 Some studies \citep[e.g.][]{Bothwell2013, Bothwell2016b, Brown2018} have explored a similar relation where the SFR is replaced by the gas content and found that, once the much lower statistics is taken into account, the resulting 3D relation is even tighter. More specifically, the anticorrelation between metallicity and gas content is tighter than the anticorrelation between metallicity and SFR, suggesting that the former is more fundamental while the latter is a by-product resulting from the dependence of the SFR on gas via the Schmidt-Kennicutt relation \citep{Schmidt1959, Kennicutt1998}. These findings further support the interpretation that the primary driver of the FMR is due to gas accretion diluting the gas-phase metallicity and fostering star formation.
 
 The existence of the FMR has been explored also on  spatially-resolved scales \citep[e.g.][]{Barrera-Ballesteros2017, Cresci2019, Sanchez-Almeida2019, Sanchez2019} by using the integral field spectroscopic data from surveys such as MaNGA \citep[][]{Bundy2015}, CALIFA \citep[][]{Sanchez2012} and SAMI \citep{Bryant2015}. Although different studies have reached somewhat different conclusions, there are indications that the FMR also holds for sub-galactic regions \citep[see discussion in ][]{Maiolino2019}.  Questions have also been raised about the evolution of the FMR with redshift \citep[e.g.][]{Mannucci2010, Troncoso2014, Sanders2018, Cresci2019, Sanders2020}. It has been shown that once the metallicity calibrations and uncertainties are taken into account the FMR does not evolve up to z$\sim$2.5 and possibly even to z$\sim$3.5-4 \citep{Cresci2019, Sanders2020}, although testing the evolution to very high redshift is made difficult by the potential evolution of the metallicity diagnostics and the limited availability of metallicity tracers with ground-based observations.
 
 The existence of the FMR has also been used to explain the redshift evolution of the mass-metallicity relation \citep{Maiolino2008,Troncoso2014,Mannucci2010,Sanders2020}, in the sense that high redshift galaxies are more star forming and therefore populate the region of the FMR with lower metallicities, resulting into an evolution of the MZR.
 
 \indent However, the past observational studies of the FMR have only focused on those galaxies classified as `star forming' according to the classical emission line diagnostic diagrams \citep[the so-called BPT diagrams][]{Baldwin1981,Kewley2006,Kauffmann2003}. This is primarily because the nebular line metallicity diagnostics (derived either empirically or through photoionization models) have mostly been calibrated on, and therefore applicable to, `star forming' galaxies classified as such based on the BPT diagarms. These diagrams provide a useful way to broadly divide the excitation mechanism of the 
 interstellar medium in galaxies in different classes, specifically: galaxies which have ongoing star formation and host young O and B stars, and the population of non-star forming galaxies as those including low-ionization (sometimes nuclear) emission region,  \citep[LI(N)ERs,][]{Belfiore2016}, with little star formation activity possibly ionized by evolved asymptotic giant branch (AGB) stars, post-AGB stars \citep[][]{Singh2013}, white dwarfs \citep[][]{Stasinska2008}, shocks \citep[][]{Heckman1980}, or radiation leaked by star forming regions, as well as active galactic nuclei (AGNs) powered by a central mass-accreting super-massive black hole. Yet, different types of galaxies are classified in these diagrams by `razor-cut' dividing lines, which are certainly over-simplistic 
 demarcations. The distribution of galaxies in these diagrams is not bimodal, but it shows a continuous distribution across the dividing lines. It is unrealistic to think that galaxies suddenly change excitation mechanism as they cross the dividing lines. Variation in properties such as nitrogen abundance, ionization parameter, age of the stellar population, can result in galaxies to spread beyond the dividing lines without necessarily implying a change in excitation mechanisms. Therefore, imposing a sharp selection based on an assumed dividing line in the BPT diagrams may introduce biases in the galactic properties and, therefore, on the inferred scaling relations with metallicity. Moreover, the nebular lines used in the BPT diagrams are also used in most metallicity diagnostics, therefore selecting galaxies through line ratios that are somehow linked with the ratios used to measure the metallicity may result in biases in the resulting metallicity measurements. Ideally, the selection of galaxies should not rely on any selection based on nebular line ratios that are used to measure the metallicity of galaxies or which can bias the properties of the selected galaxy population.
 
 An additional problem in selecting only star forming galaxies based on their BPT classification is that it limits our capability of fully exploring the metallicity scaling relations even beyond `classical' star forming galaxies on the BPT diagram. When galaxies decrease their star formation and move towards quiescence, they generally still have plenty of gas \citep{Gobat2018}, but which has no longer the emission line ratios of star forming galaxies. When detected in nebular emission lines, quiescent or (many of) `green valley' galaxies have generally LI(N)ER-like emission line ratios \citep{Belfiore2018}, which are typically associated with Diffuse Ionized Gas (DIG) excited either by evolved star (e.g. hot post-AGB stars), or by hardened, leaked radiation from residual star forming regions, as discussed above. However, as the metallicity diagnostics for galaxies in this region of the BPT diagram have not been calibrated until recently, these important stages of galaxy evolution have not been included in the metallicity scaling relations. 
 Chemical evolution studies of non star-forming  galaxies have been either restricted to the use of stellar metallicities \citep[][]{Gallazzi2014, Peng2015, Spitoni2017, Trussler2020, Trussler2021}, analytical models \citep[][]{Spitoni2017} and cosmological simulations \citep[][]{RomeoVelona2013} where gas-phase metallicity measurements are not needed.
 It would be important to investigate the same process by probing the gas phase metallicites in green valley and passive galaxies.

 \indent Hence, the main caveat in addressing the various issues outlined above, i.e. for extending FMR to  galaxies classified as non-star forming in the BPT diagrams, has been the lack of suitable metallicity calibrators for estimating the gas-phase metallicity of such systems. In order to overcome this problem, \citet{Kumari2019} utilised the IFS data of 24 nearby star-forming galaxies taken with the MUSE instrument on the Very Large Telescope (VLT) to identify and correct biases in the metallicity measurements of low ionization emission regions (LIERs), or regions dominated by DIG. The method consisted in comparing the emission line properties of LIER/DIG regions to those of nearby H \textsc{ii} regions with similar levels of chemical enrichment, thus devising novel calibrations for the DIG/LIERs. Since the emission line ratios of DIG/LIERs fall on the same region occupied by the galaxy populations classified as non-SF in the BPT diagrams, the DIG/LIER calibrators are also applicable to the non-star-forming/passive/quiescent galaxy populations, and hence enable measurements of their gas-phase metallicities (see discussion in \citealt{Kumari2019}). 
 
\begin{figure}
 	\centering
 	\includegraphics[width=0.45\textwidth]{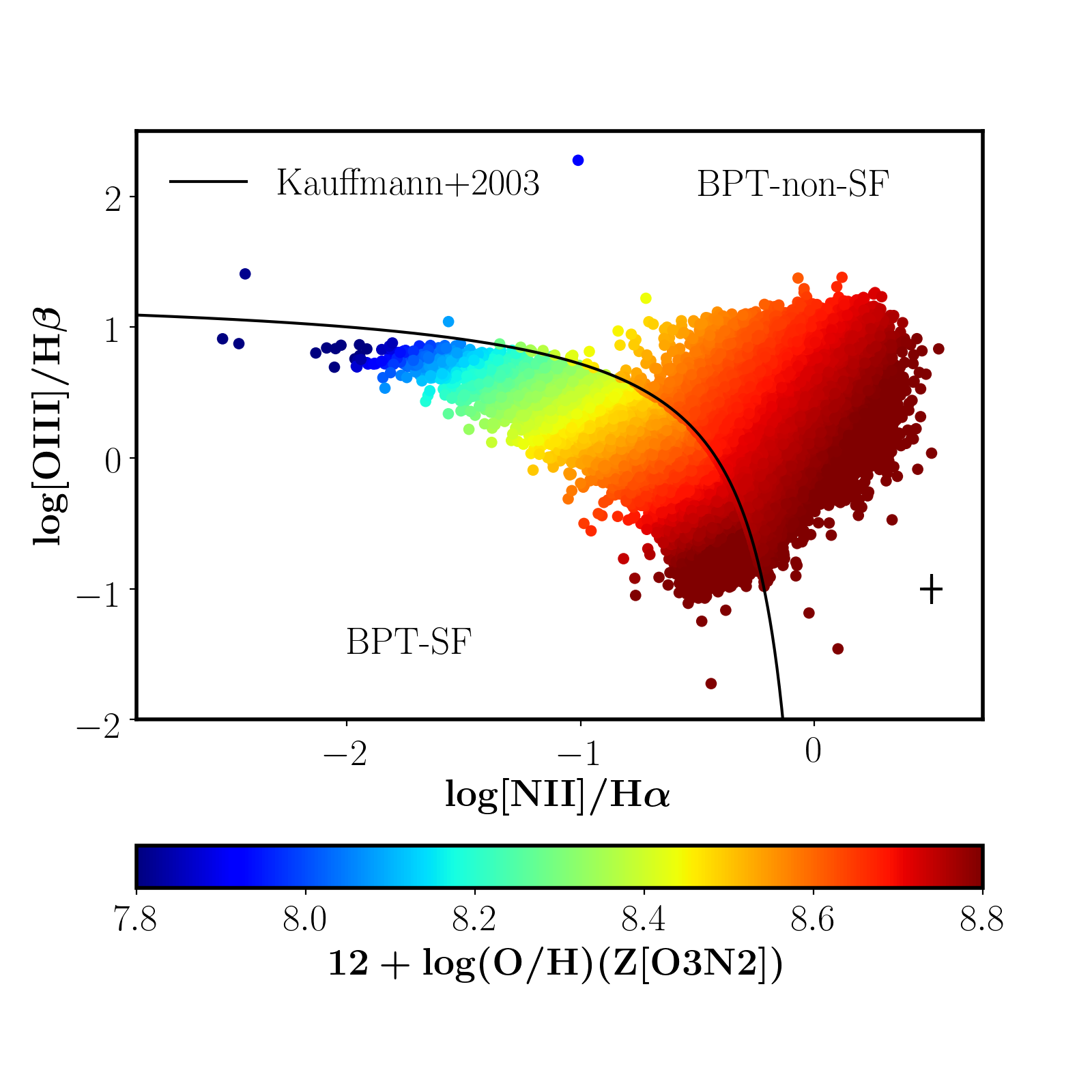}
 	\includegraphics[width=0.45\textwidth]{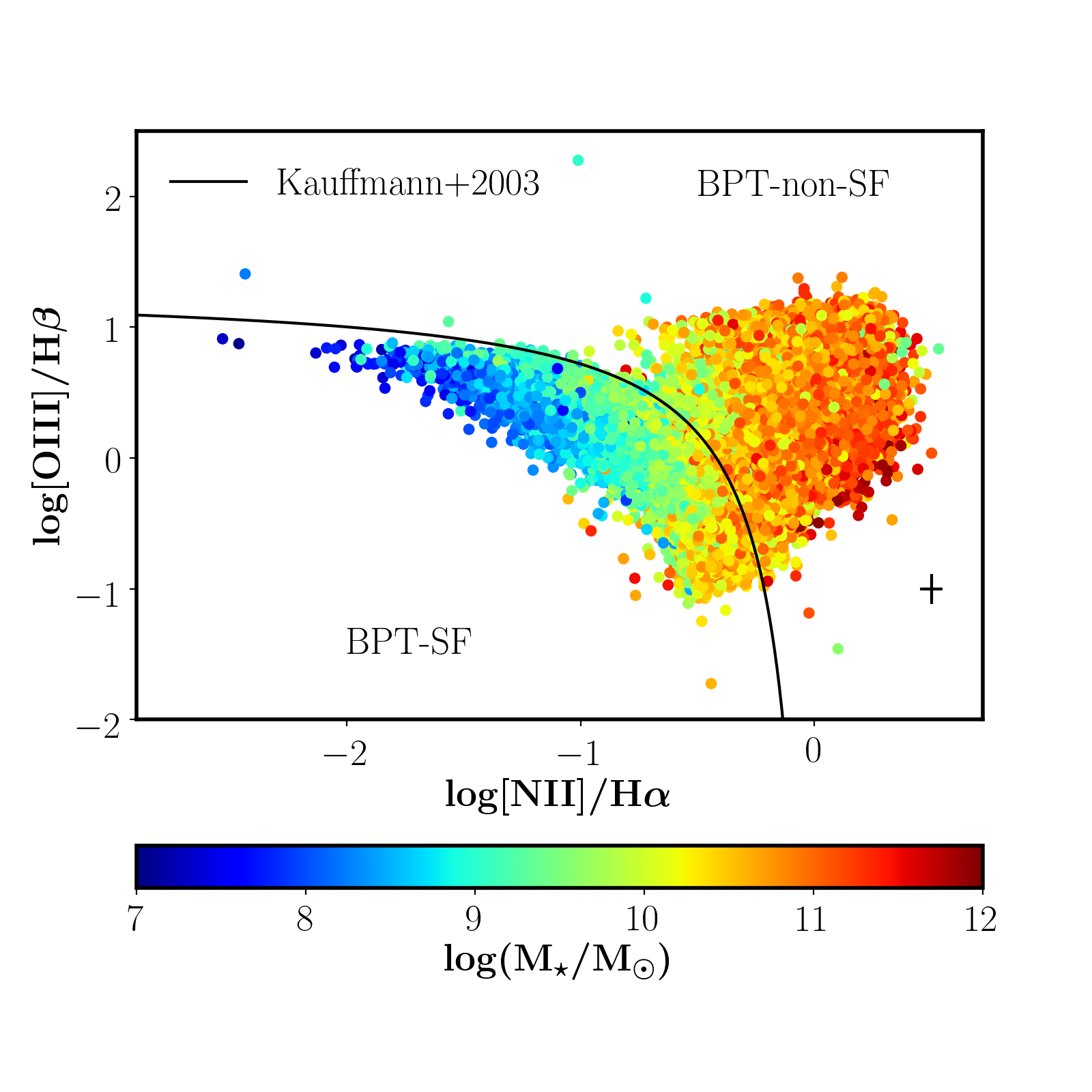}
 	\caption{[O \textsc{iii}]/H$\beta$ versus [N \textsc{ii}]/H$\alpha$ diagram.  In the upper panel, the data-points are colour-coded with respect to the gas-phase metallicity (12 + log(O/H)) using the O3N2 diagnostic. We use recipes given in \citet{Curti2017} and \citet{Kumari2019} to estimate gas-phase metallicities of BPT-SF and BPT-non-SF galaxies, respectively. In the lower panel, the data points are color-coded with respect to stellar mass. On both panels, the solid black curve corresponds to the line dividing star forming galaxies from non-star forming galaxies  empirically derived by \citet{Kauffmann2003}, where the galaxies lying below and beyond solid black curves are referred to as BPT-SF and BPT-non-SF galaxies, respectively. The error bars in the bottom-right corner shows the uncertainty on the emission line ratios.}
 	\label{figure:BPT}
 \end{figure}

 
 \indent The aim of this paper is to extend the FMR to the massive emission line galaxies lying beyond the star-forming sequence on the BPT diagrams. Henceforth, we use the terms "BPT-SF" for galaxies lying below the dividing star-forming line on the classical BPT diagrams (Figure \ref{figure:BPT}, while "BPT-non-SF" for galaxies lying beyond the dividing star forming line, occupied by the AGN/LI(N)ERs, as explained in more detail in Section \ref{section:sample params}).  We mainly focus on emission line galaxies because the gas-phase metallicities of these systems can be estimated from the recent prescriptions from \citet{Kumari2019}, based on the strong emission lines H$\beta$, [O \textsc{iii}]$\lambda$5007, H$\alpha$ and [N~\textsc{ii}]$\lambda$6584. For this purpose, we use the vast dataset of SDSS-DR7 available from the Max Planck for Astrophysics and Johns Hopkins University groups (MPA-JHU) which has successfully been used for such studies in the past. We not only study the conventional form of FMR, (i.e. the 3D relation between  M$_{\star}$, SFR and 12 + log(O/H)), but we also build upon the previous work of \citet{Peng2015} and \citet{Trussler2020} to explore the difference of gas-phase metallicities of green valley and quiescent galaxies and the Main Sequence (MS) galaxies while considering the offset in their SFR activity from the Main Sequence population, in order to investigate their chemical composition and processes that might lead to quenching.

 \indent The paper is organized as follows. Section \ref{section:sample params} describes the SDSS dataset and the selection criteria, along with the methods to estimate the gas-phase and stellar metallicites, stellar masses and star-formation rates of BPT-SF and BPT-non-SF galaxies in the sample. Section \ref{section:FMR extension} presents the FMR for  BPT-SF sample, BPT-non-SF sample and the combined sample of BPT-SF and BPT-non-SF samples, where we use the [N~\textsc{ii}]-BPT and [S~\textsc{ii}]-BPT diagnostic diagrams to classify the SF and non-SF galaxies.  We also study an alternative representation of the FMR for the combined sample of BPT-SF and BPT-non-SF galaxies for an easier interpretation. Section \ref{section:discussion} discusses and summarises our main results. 

 \section{Sample selection and measurement of FMR parameters}
 \label{section:sample params}
 \subsection{Galaxy sample}
 \label{section:sample}
 \indent We use the SDSS-DR7 MPA-JHU catalogue\footnote{Available at  \url{http://www.mpa-garching.mpg.de/SDSS/DR7}}, which provides the emission line fluxes, star-formation rates and stellar masses. As we describe in Section \ref{section:gas Z}, our analysis is dependent on the use of the [N~\textsc{ii}]-BPT diagnostic diagram for applying the metallicity calibrations, as this is the diagram used by \cite{Curti2017} to select SF galaxies for their empirical calibration, which has then been extended by \cite{Kumari2019} to the non-SF regions. However, we note that the results do not change significantly if the classification of galaxies (used to apply the metallicity calibrations) is based on the [SII]-BPT diagram (the results based on this classification are shown in the appendix \ref{appendix:noagn}). We apply 
 a S/N cut of 3 on the H$\alpha$ and H$\beta$ lines and a S/N cut of 2 on [O \textsc{iii}] $\lambda$5007 and on 
 [N \textsc{ii}] $\lambda$6584. Such a S/N cut ensures the clean BPT diagnostic diagrams with no spurious line ratios. We have verified that the results do not change significantly by increasing the S/N thresholds, simply resulting in lower statistics. In addition to S/N cuts, we also remove all galaxies whose photometric flags consists of DEBLEND\_NOPEAK
 or DEBLEND\_AT\_EDGE, and restrict the sample to galaxies with total stellar masses in the range 10$^{7-13}$ M$_{\odot}$ thus ensuring the inclusion of low-mass dwarfs as well as massive galaxies. Different studies have assumed different redshift cuts in the previous FMR studies. We found that a lower-limit on redshift do not affect the higher-mass end of the FMR, which is the regime we want to probe in this work.  Still, we have imposed a redshift cut of z>0.016, which ensures that the SDSS fibres cover  at least $\sim$1 kpc on the selected galaxies. This results in a sample of $\sim$10$^5$ galaxies. 

 
 \subsection{Gas-phase metallicity}
 \label{section:gas Z}
 \indent We aim to extend the FMR study to the BPT-non-SF galaxies which requires their metallicity calibration. As already discussed, \citet{Kumari2019} provides the first metallicity calibrations for the DIG/LIER and for the BPT-non-SF galaxies. These empirical calibrations for BPT-non-SF galaxies are based on two distinct diagnostics, O3N2 (= log([O~\textsc{iii}]/H$\beta$) - log([N~\textsc{ii}]/H$\alpha$)) and O3S2~(=~log([O~\textsc{iii}]/H$\beta$~+~ [S~\textsc{ii}]/H$\alpha$)). However, the O3S2 values exhibited by the full SDSS sample selected here lies within $\sim$ $-$0.4-1.6, which is beyond the range in which the O3S2 diagnostic has been calibrated ($\sim$ $-$0.5--0.8, see e.g. \citet{Maiolino2019}). Hence, we only use metallicities derived from the O3N2 diagnostic for the entire analysis.
 
 
 
 \indent Figure \ref{figure:BPT} (upper-panel) shows the [N \textsc{ii}]-BPT diagnostic diagram (i.e. [O \textsc{iii}]/H$\beta$ versus [N \textsc{ii}]/H$\alpha$) along with the empirical starburst line from \citet{Kauffmann2003}, which is used to classify the SF and non-SF galaxies.   We do not remove the AGNs from non-SF sample as such a recipe is not available for [N \textsc{ii}]-BPT  diagram.  Moreover, \cite{Kumari2019} make the case that their new metallicity calibration also extend to the AGN/Sy population (thanks to the [OIII]/H$\beta$ corrective term mentioned below), though further tests on AGN needs to be done. However, in  Appendix \ref{appendix:noagn} we show that the inclusion or removal of AGNs (based on the [SII]-BPT classification) hardly affects the overall results.

In Figure \ref{figure:BPT} (upper-panel), BPT-SF and BPT-non-SF samples are colour-coded with respect to their gas-phase metallicities estimated by using the empirical prescriptions given in \citet{Curti2017} and \citet{Kumari2019}, and can also be expressed as the following equations.
 \begin{equation}
\rm 12+log(O/H)_{BPT-SF} = 7.64 + 0.22 \times \sqrt{25.25-9.072\times O3N2},
 \label{eq: oh sf}
 \end{equation}
 
 \begin{multline}
 \rm 12+log(O/H)_{BPT-non-SF} = 7.673 \\
 + 0.22 \times \sqrt{25.25-9.072\times O3N2} + 0.127\times O3,
 	\label{eq: oh nonsf}
 \end{multline}
 \noindent where O3 = log([O \textsc{iii}]/H$\beta$).

 \subsection{Stellar metallicity}
 \label{section: stellar Z}
 \indent We use the gas-phase metallicities of BPT-SF and BPT-non-SF galaxies for the majority of our analysis. However, we will also compare these results with stellar metallicities of the Main Sequence, green valley and passive galaxies sample taken from \citet{Trussler2020}. Briefly, a $\chi^2$-minimisation code \textsc{firefly} \citep{Wilkinson2017} is used to fit the input galaxy spectra using a combination of stellar population models of \citet{Maraston2011}, stellar spectra from the \textsc{miles} \citep{Sanchez-Blazquez2006} and stellar libraries assuming a Kroupa initial mass function \citep{Kroupa2001}, and hence mass-weighted stellar metallicities are obtained. 
 
 \subsection{Stellar mass}
 \label{section: stellar mass}
 \indent We use the total stellar masses of our galaxy sample from the MPA-JHU catalogue. They are obtained by using fits to the total photometric magnitudes of these galaxies and are different than the method based on spectral-indices adopted by \citet{Kauffmann2003b} or \citet{Gallazzi2005} though these differences are quite small (a median offset of $-$0.01 dex between mass estimates from spectral indices and photometric magnitudes\footnote{\url{https://www.mpa.mpa-garching.mpg.de/SDSS/DR7/mass_comp.html}}). We have restricted our sample to a stellar mass-range of 10$^{7-13}$ M$_{\odot}$.
 
 The lower panel of Figure \ref{figure:BPT} shows the distribution of galaxies in our sample on the [NII]-BPT diagram, color-coded by their stellar mass. The figure clearly shows that studies selecting only BPT-SF galaxies miss a large population of massive galaxies.

 \subsection{Star-formation rate}
 \label{section: SFR}
 \indent We obtain the SFR of all galaxies including the SF, LINERs, AGNs and composite galaxies from the MPA-JHU catalogue. These SFR estimates are obtained by fitting the models of \citet{Charlot2001} to the SF galaxies as discussed in  \citet{Brinchmann2004} with slight modifications \footnote{https://wwwmpa.mpa-garching.mpg.de/SDSS/DR7/sfrs.html}, for LINERs, AGNs, composites and for low S/N SF galaxies. Some of these modifications correspond to estimating the average conversion factor from observed H$\alpha$ luminosity to SFR for low S/N galaxies, different dust-attenuations and appropriate aperture corrections for galaxies with low level of star-formation following \citet{Salim2007}.
 
 \indent We use the SFR estimates from MPA-JHU catalogue mainly because it provides the SFR within the 3 arcsec aperture of the SDSS fibre. The other SFR estimates of the SDSS sample \citep[e.g.][]{Salim2016, Bourne2016, Oemler2017} may provide an improvement over the total SFR estimates from the MPA-JHU catalogue but are not suitable for the purposes of this work. Since the eventual aim of our work is to study the correlation between SFR and gas-phase metallicity, it is more appropriate to study these quantities within the same spatial region. In particular, gas accretion resulting into metallicity dilution and star formation boosting can only be properly probed by measuring metallicity and SFR in the same aperture. Given that we only have information on the gas-phase metallicity within the galactic region covered by the fibre, we prefer to use the fibre SFR. Stellar masses, on the other hand, acts more like a clock telling us about the evolution stage of a galaxy (see Section \ref{section:discussion}) as well as a tracer of the overall gravitational potential well of the galaxy, therefore we used the total M$_{\star}$ in this work. 
 
 \indent We note that a few FMR studies \citep[e.g.][]{Mannucci2010} have estimated SFRs from H$\alpha$ luminosities following the prescription of \citet{Kennicutt1998}. However, this calibration is only suitable for SF galaxies or H \textsc{ii} regions as there is an inherent assumption of photoionization by hot O and B stars, rather than other more evolved stellar populations (or even AGNs) which are likely the source of ionization within non-SF galaxies.  
 
 \begin{figure}
 	\centering 
 	\includegraphics[width=0.45\textwidth]{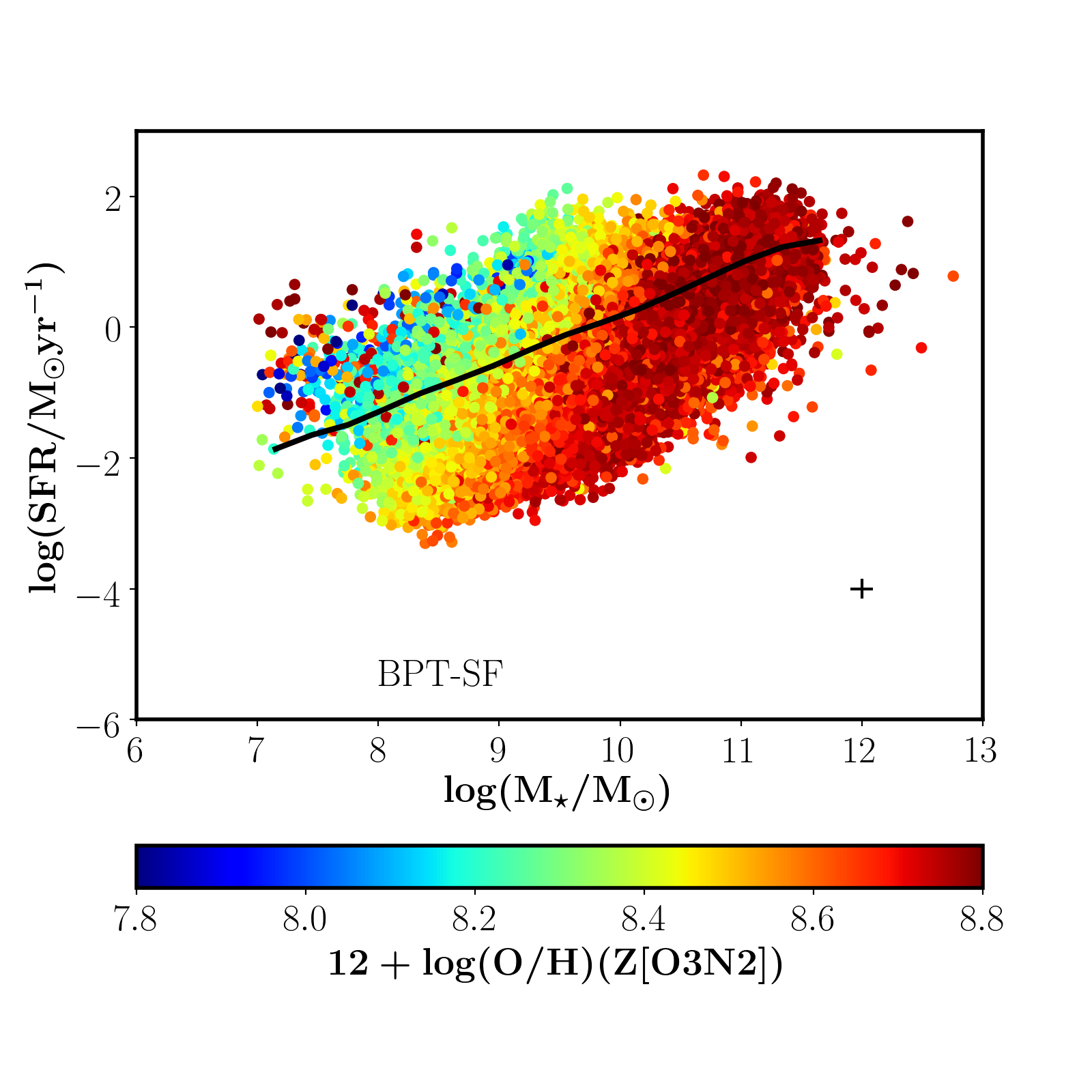}
 	\includegraphics[width=0.45\textwidth]{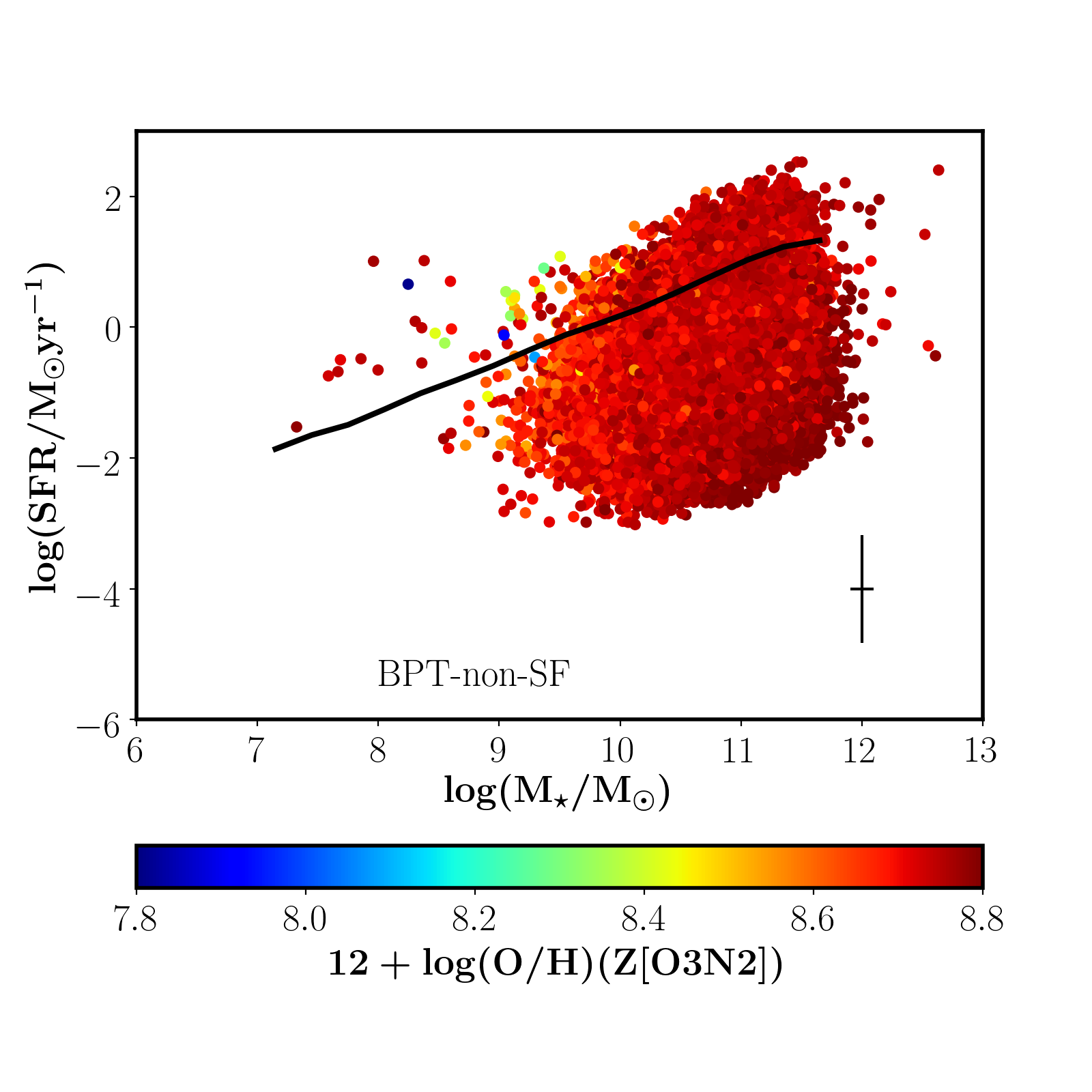}
 	\caption{SFR versus M$_{\star}$ for the  BPT-SF (upper-panel) and BPT-non-SF (lower-panel) samples identified on the basis of [N \textsc{ii}]-BPT. The data points are colour-coded with respect to the gas-phase metallicity (12 + log(O/H)) using the O3N2 diagnostic described in Section \ref{section:gas Z}. On both panels, the black Main sequence line is derived from fitting a median to the SF galaxies. The errorbars in the bottom-right corner indicate the median uncertainties on SFR and M$_{\star}$ within each subsample and are estimated from the percentiles provided in the MPA-JHU catalogue. The uncertainty on gas-phase metallicity is of the order of 0.09 dex.} 
 	\label{figure:MS samples}
 \end{figure}
 
 \indent Figure \ref{figure:MS samples} shows SFR versus M$_{\star}$ for the  BPT-SF (upper-panel) and BPT-non-SF (lower-panel) samples identified on the basis of the [N \textsc{ii}]-BPT selection. The solid black line shows the location of the star formation Main Sequence. The latter is derived by taking the median of the SFRs of the BPT-SF galaxies in the stellar mass bins of 0.3 dex. Note that different definitions of  Main Sequence have been proposed by various groups, \citep[e.g.][]{RenziniPeng2015, Popesso2019} which are valid in different stellar mass ranges and generally fitted by a linear form, but may show bending depending on the SFR recipes adopted. However, these other studies use the  {\it total} SFR and stellar mass, hence are not suitable for this work which uses the star-formation activity within the spatial region (3 arcsec SDSS fibre) and the related metallicity enrichment/dilution.
 
 Figure \ref{figure:MS samples} shows, not surprisingly, that the BPT-non-SF galaxies have SFRs typically below the SF Main Sequence, even by up to 3 dex lower, so they are quite suitable for our goal of extending the FMR to galaxies approaching quiescence.
 
  About 10$\%$ of BPT-non-SF galaxies have SFRs higher than the Main Sequence. This is  in agreement with the results of \citet{Leslie2016}, who find that there are galaxies classified as ``composite'' or ``ambiguous'' in the SDSS sample which lie on or above the star-forming Main Sequence. These results are also in line with
 \citet{Oemler2017}, who report a large population of quiescent galaxies with disproportionately high SFRs.
 The reason why these highly star forming galaxies are not classified as SF in the BPT-diagram may be a consequence of heavy obscuration (which leaves only the outer, LI(N)ER like regions visible), the presence of shocks, or nitrogen enrichment, and highlights how the simple BPT classification may be inadequate and that it can result into seriously biasing the sample. However, we find that only $\sim$2$\%$ of BPT-non-SF galaxies are massive ((log(M$_{\star}$/M$_{\odot}$) $>$ 11)  and have SFRs above the Main Sequence, thus ensuring that FMR results presented here are not biased by this galaxy population.
 
 
 \section{Results: extending the FMR to BPT-non-SF galaxies}
 \label{section:FMR extension}
 
 \begin{figure*}
 	\centering
 	\includegraphics[width=0.45\textwidth]{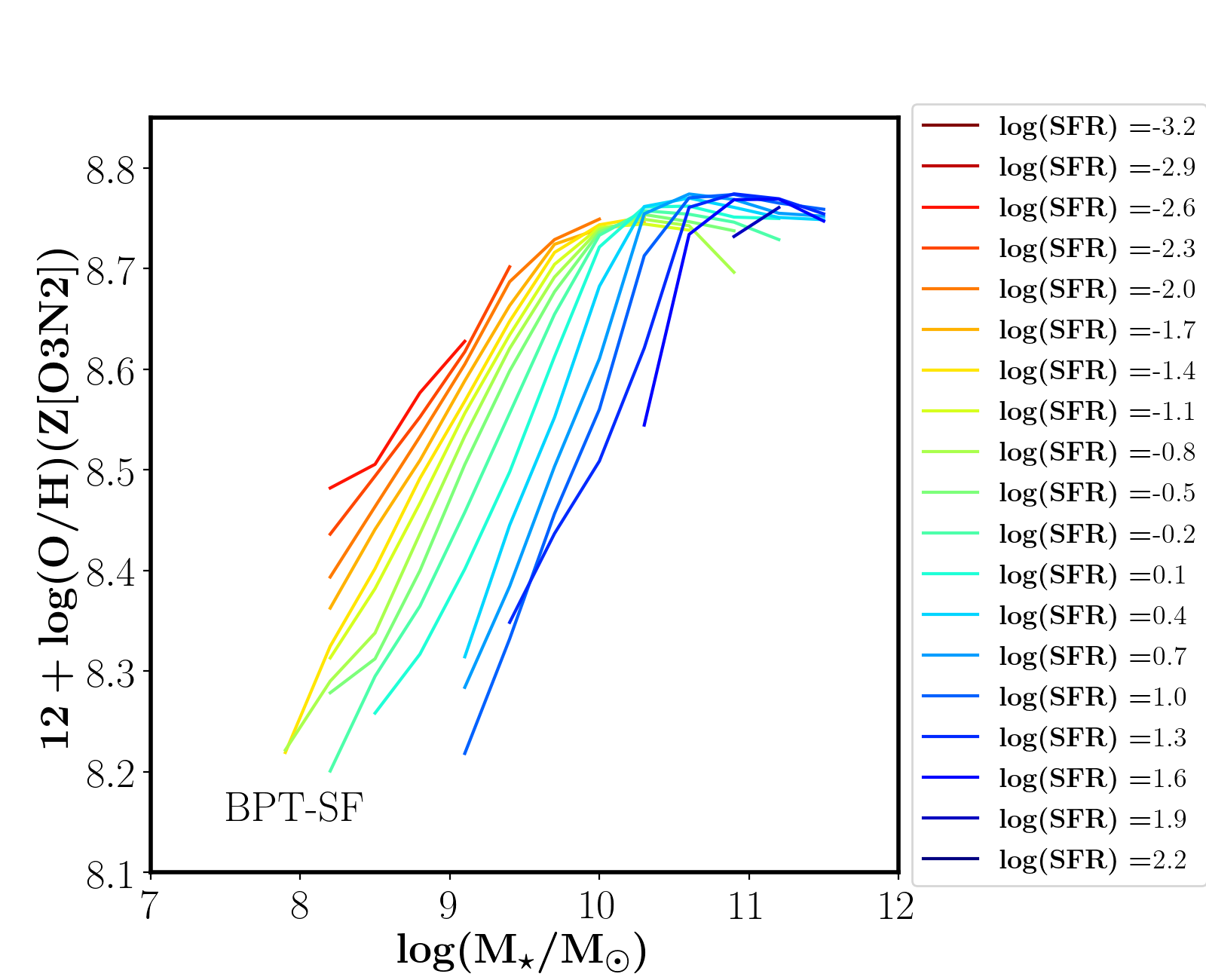}
 	\includegraphics[width=0.45\textwidth]{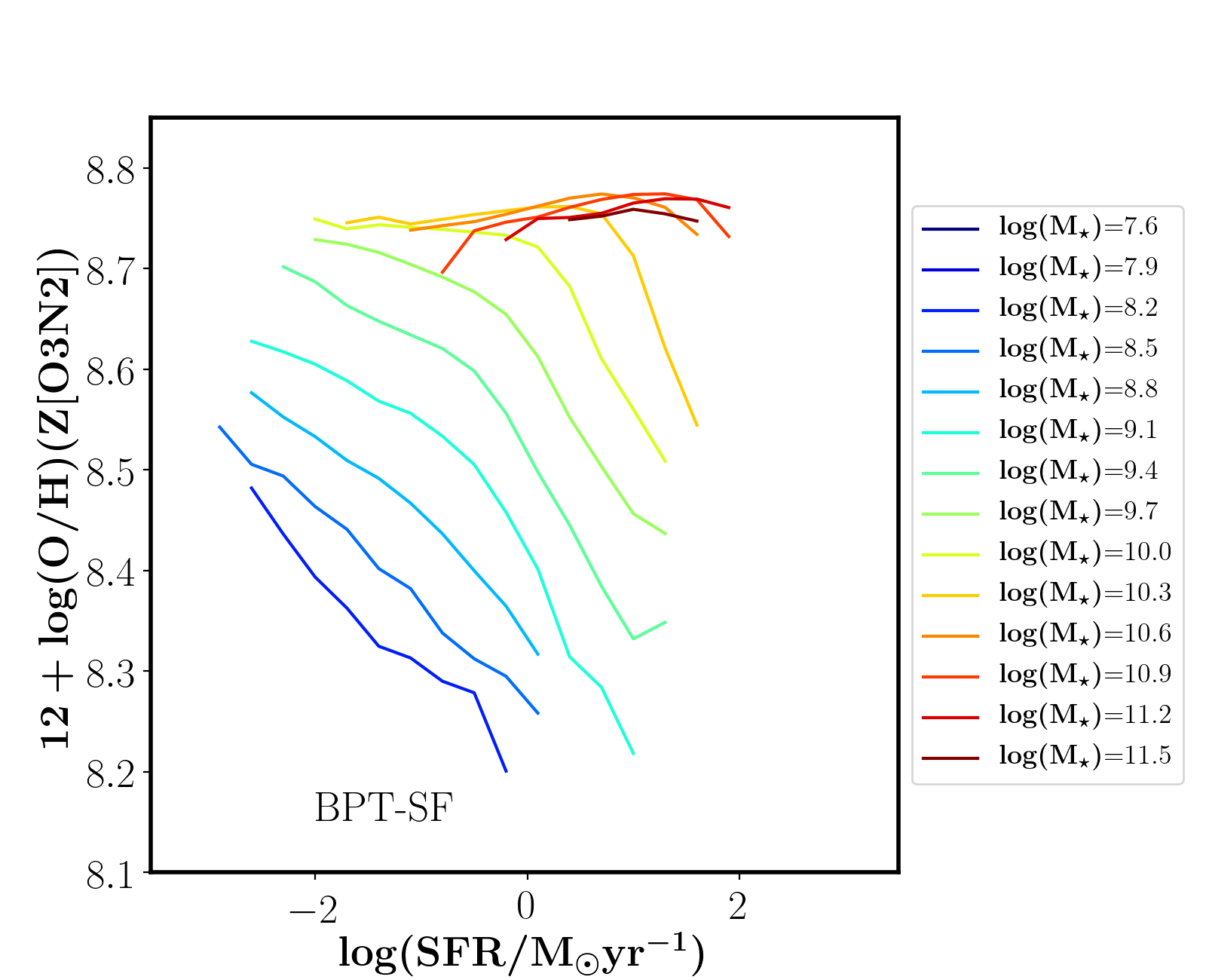}
 	
 	\centering
 	\includegraphics[width=0.45\textwidth]{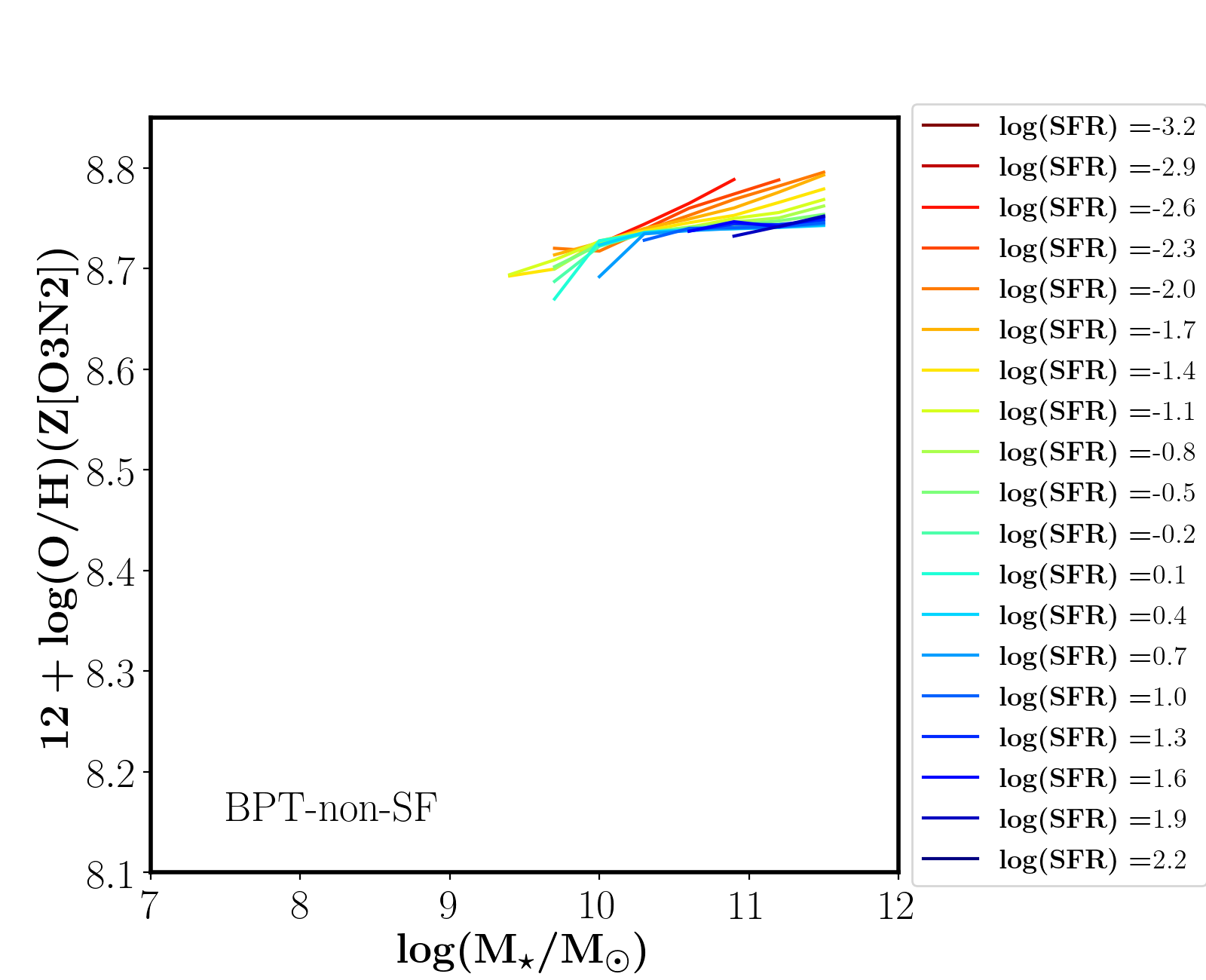}
 	\includegraphics[width=0.45\textwidth]{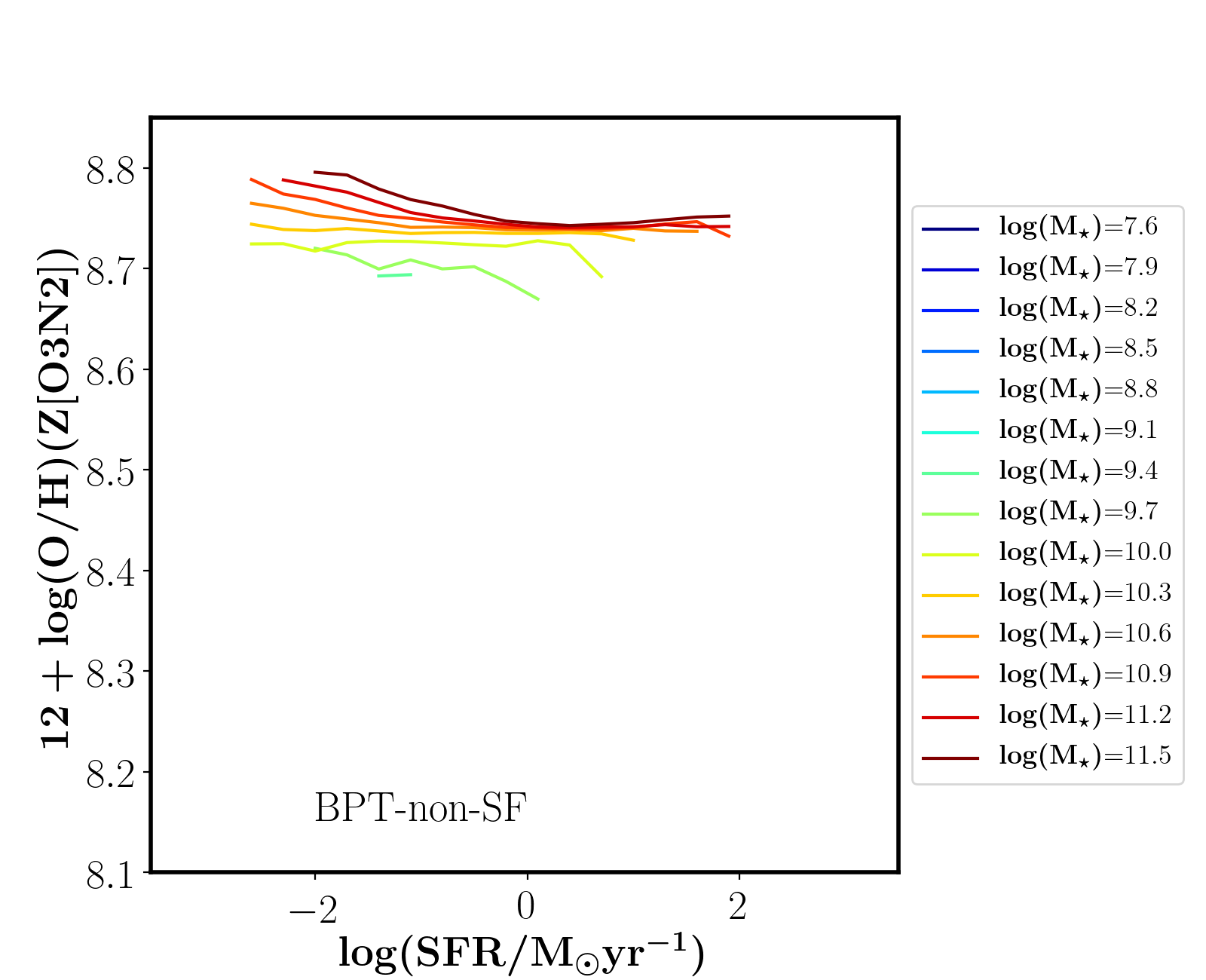}
 	
 	\centering
 	\includegraphics[width=0.45\textwidth]{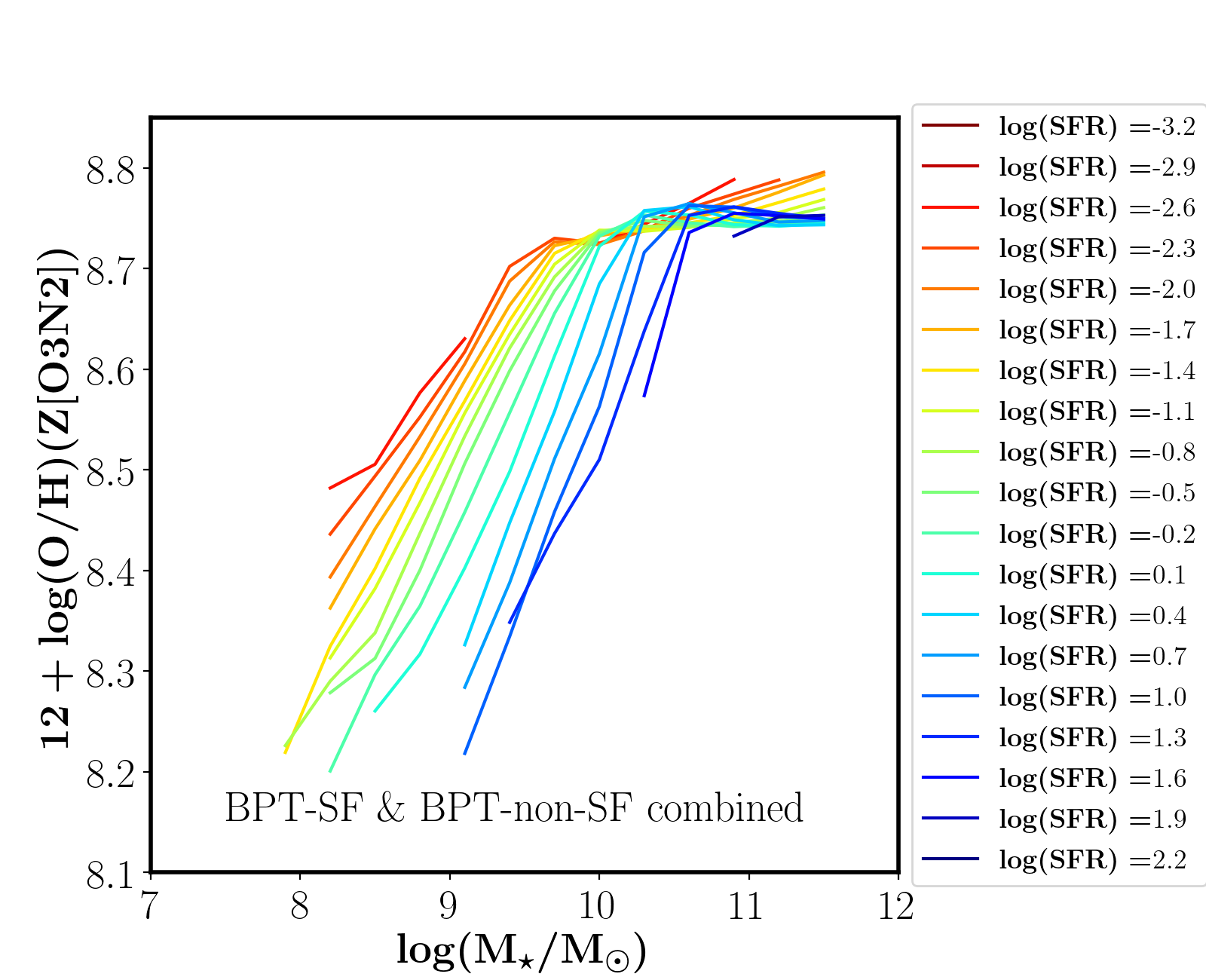}
 	\includegraphics[width=0.45\textwidth]{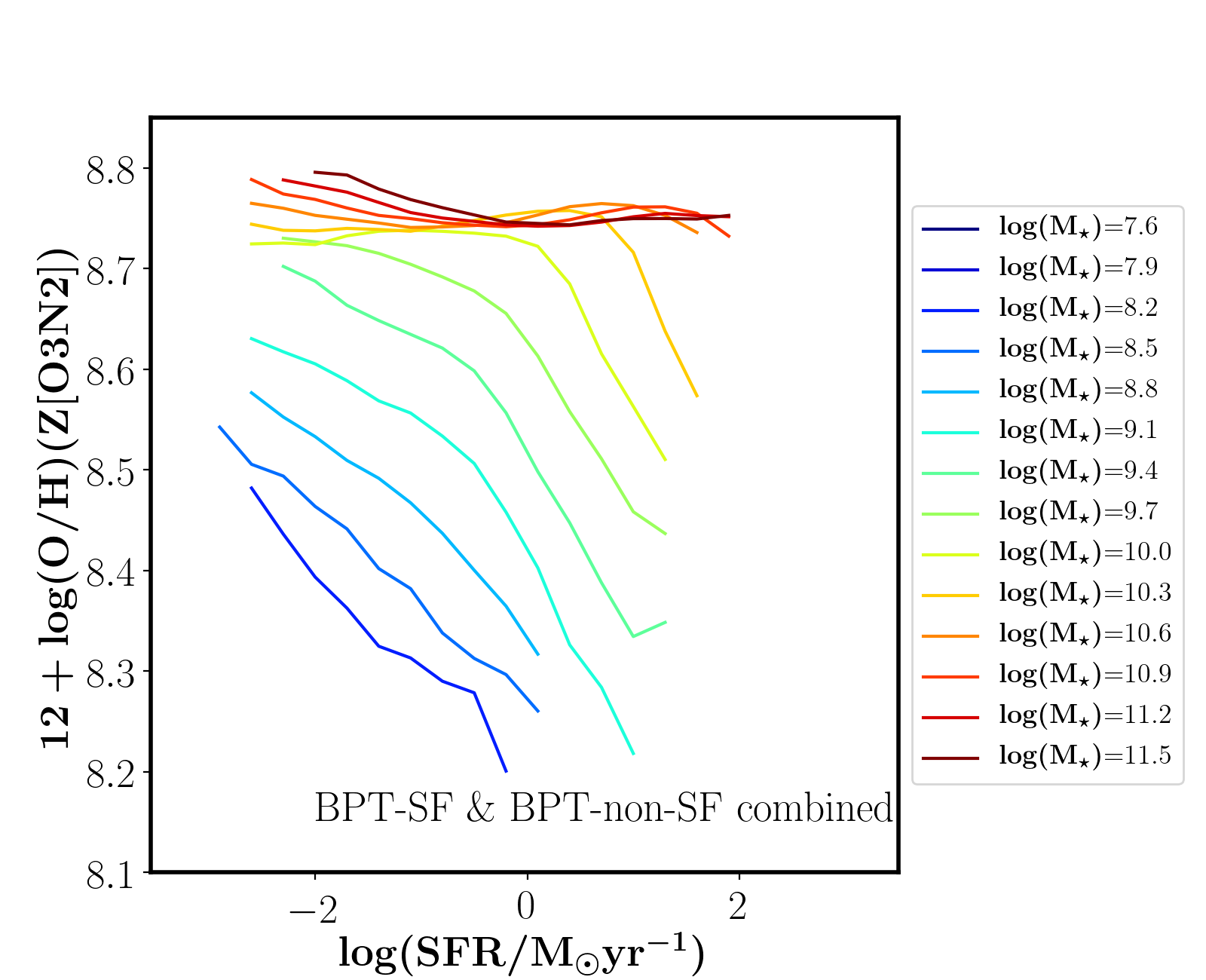}
 	
 	\caption{Left-hand panel: The gas-phase metallicity (12 + log(O/H)) versus stellar mass (log M$_{\star}$/M$_{\odot}$) for BPT-SF galaxies (upper-panel), BPT-non-SF galaxies (middle panel) and their combined sample (bottom panel). The coloured curves on each plot in the left-hand panel correspond to median metallicites estimated in the bins of log M$_{\star}$/M$_{\odot}$ and star-formation rate (log SFR) and color-coded with respect to log SFR in a given bin. Right-hand panel: The gas-phase metallicity (12 + log(O/H)) versus log SFR for  BPT-SF (upper-panel), BPT-non-SF (middle panel) and their combined sample (bottom panel). The coloured curves on each plot in the right-hand panel correspond to the median metallicities estimated in bins of log SFR and log M$_{\star}$/M$_{\odot}$. The [N \textsc{ii}]-BPT classification along with the maximum star-burst line from \citet{Kauffmann2003} is used to separate SF (upper-panel) and non-SF galaxies (middle panel). The gas-phase metallicities of BPT-SF (upper-panel) and BPT-non-SF (middle panel) galaxies are obtained using the empirical calibrations involving the O3N2 diagnostic provided in \citet{Curti2017} and \citet{Kumari2019}, respectively. The bottom panel consisting of both BPT-SF and BPT-non-SF galaxies and hence their metallicities are obtained from both SF \citep{Curti2017} and non-SF \citep{Kumari2019} calibrations, depending on the type of galaxy populations.} 
 	\label{figure:conventional FMR}
 \end{figure*}

 \begin{figure*} 
 	\centering
 	\includegraphics[width=0.75\textwidth]{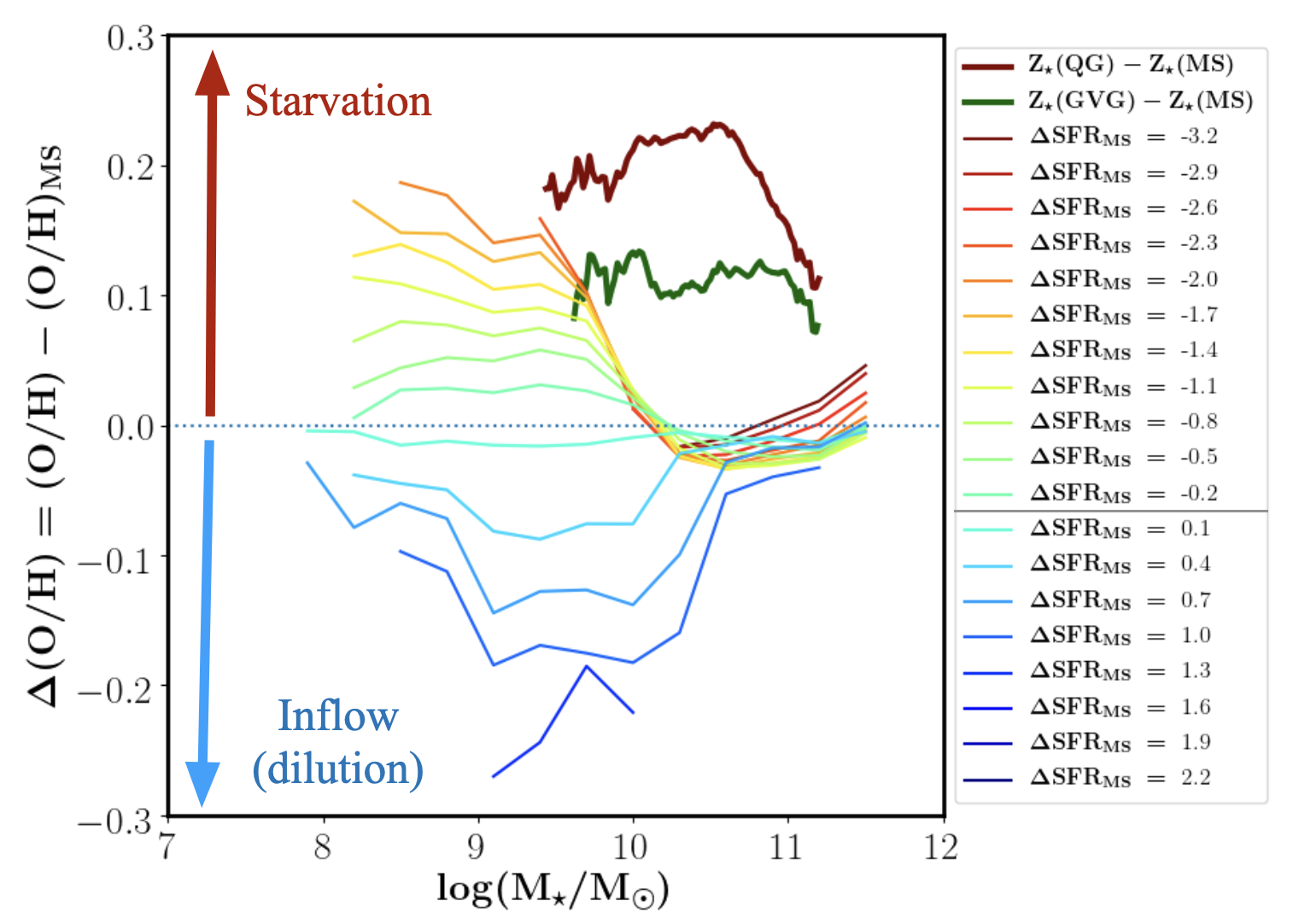}
 	\caption{A comparison of metallicity difference versus stellar mass for the gas-phase metallicity as well as for stellar metallicity for the SDSS galaxies. The gas-phase metallicity difference $\rm \Delta (O/H)$ with respect to that of MS galaxies are estimated for the combined sample of BPT-SF and BPT-non-SF galaxies, identified using the [N \textsc{ii}]-BPT and are shown by thinner coloured curves corresponding to $\rm \Delta SFR_{MS}$ varying from -3.2 to 2.2. The thick maroon curve corresponds to the stellar  metallicity difference between quiescent galaxies and MS galaxies (Z$_{\star}$(QG) - Z$_{\star}$(MS), while the thick dark green curve corresponds to the stellar metallicity difference between green valley galaxies and MS galaxies (Z$_{\star}$(GVG) - Z$_{\star}$(MS). The stellar metallicity differences are taken from \citet{Trussler2020}.} 
 	\label{figure:master plot}
 \end{figure*}

 \subsection{Conventional FMR}
 \label{section:conventional}
 \indent Figure \ref{figure:conventional FMR} shows the two projections of the conventional form of the FMR, i.e. the relation between stellar mass, gas-phase metallicity and star-formation rate. 
 We estimated the median gas-phase metallicities in the bins of stellar mass and SFR of 0.3 dex and we retain only those which have at least 30 galaxies in each bin. The left-hand panels (Figure \ref{figure:conventional FMR}) show the mass versus metallicity colour-coded with respect to the SFR, while the right-hand panels (Figure \ref{figure:conventional FMR}) show the SFR versus metallicity colour-coded with respect to stellar mass, for the three samples as described below. 
 
 \begin{itemize}
 	\item\textit{BPT-SF Galaxies}  (Figure \ref{figure:conventional FMR}, Upper panel): The left-hand panel shows that at low-stellar masses, the SF galaxies with higher SFR have lower metallicity while at high-stellar masses, there appears to be a small inversion ($\sim$0.05 dex) in this trend, i.e. metallicity increases with SFR. The same systematic dependence of metallicity on SFR for different stellar mass is shown in the another projection of the FMR (right-hand panel), where gas-phase metallicity is plotted with respect SFR and each curve is colour-coded with respect to the log M$_{\star}$. 
 	
 	\indent The inversion at the high-mass end, though quite small, aligns with previous studies \citep[e.g.][]{Yates2012, Lara-Lopez2013}, but is in contrast to the results of \citet{Mannucci2010}, the first FMR study of SF galaxy which instead shows a flattening of the gas-phase metallicity at high stellar-mass. Such a difference is explained by a combined effect of  different S/N cuts and different methods to estimate SFR and metallicities \citep{Cresci2019}. We reproduce  the results of \citet{Mannucci2010} if we perform a S/N cut of 25 on H$\alpha$ emission line as done in their work. Such a high S/N cut on H$\alpha$ emission line is needed to avoid introducing metallicity biases in the lower-metallicity regime, where [N \textsc{ii}] is siginificantly fainter than H$\alpha$, but this S/N cut only includes SF galaxies with very high SFR, that defies the purpose of this work focussing on the extension of the FMR to BPT-non-SF galaxies and must also include the galaxies with very low SFRs. 
 	Our relaxed requirement on the S/N on H$\alpha$ and H$\beta$ enables us to explore lower levels of SFR; more specifically, the minimum SFR bin in our work corresponds to log(SFR/M$\rm_{\odot}yr^{-1}$) $\sim -3.2$ which is significantly lower than $-$1.45 of \citet{Mannucci2010}. The S/N cut on the [NII] line may potentially introduce a bias in the metallicity distribution, however, as we will discuss, this would make our findings even stronger. Anyhow, since the main purpose of this work is to extend the FMR into the LI(N)ER region (next section) where the [NII] emission is as strong as H$\alpha$, we do not expect that the S/N cut on [NII] introduces a significant bias. 
 	
 	\item\textit{BPT-non-SF Galaxies} (Figure \ref{figure:conventional FMR}, Middle panel): The left-hand panel shows that the gas-phase metallicity for BPT-non-SF galaxies increases with decreasing SFR for a given stellar mass lying in a range of log(M$\rm_{\star}/M_{\odot}) \sim 10-11.5$. Similarly, the right-hand panel shows that the gas-phase metallicity of BPT-non-SF galaxies appears to be constant at high log SFR/M$\rm_{\odot}$yr$^{-1}$ ($>$ 1.0) for a mass range spanning an order magnitude, though we find that the gas-phase metallicity increases with increasing stellar-masses for lower SFRs and that the anticorrelation of metallicity with SFR is observed primarily at log SFR/M$\rm_{\odot}$yr$^{-1}$ ($<$ 1.0).
 	
 	\indent The high SFRs probed by this sub-sample is a bit counter-intuitive for galaxies classified as non-SF galaxies, however, as already mentioned, it is consistent with other works, such as \citet{Oemler2017}, who report a large population of quiescent galaxies with disproportionately high SFRs. Such high SFRs for a non-SF classified sample, comprising LINERs and AGNs, are also consistent with the result of \citet{Leslie2016}. Yet, this classification has allowed us to probe, as already discussed, the massive BPT-non-SF galaxies whose SFRs are  $\sim$1.5 dex below the minimum SFR values probed by galaxies classified as SF (see Figure \ref{figure:MS samples}), and (as already discussed) 3 dex below the Main Sequence, and thus enables us to probe a relatively quiescent population.
 	
 	\item\textit{Combined sample of BPT-SF and BPT-non-SF galaxies}  (Figure \ref{figure:conventional FMR}, Bottom panel):
 	The left-hand panel shows that the mass-metallicity relation for the combined sample of BPT-SF and BPT-non-SF galaxies at the high-mass end is dominated by the BPT-non-SF galaxies which results in a disappearance of inversion of curves seen for BPT-SF galaxies, suggesting that the latter was primarily a result of biases introduced by the BPT selection. Instead we find that in the global (unbiased) sample at high masses the metallicity tends to flatten with stellar mass for high SFR, but then starts to increase with decreasing SFR at log SFR/M$\rm_{\odot}$yr$^{-1}$ ($<$ 0.0). These trends are also visible in the right-hand panel showing metallicity versus SFR where curves are colour-coded with respect to stellar mass.
 \end{itemize}
 
 
 
 \subsection{Alternative representation of the FMR}
 \label{section: new form}
 
 \indent  Figure \ref{figure:master plot} shows an alternative representation of the FMR, were the absolute metallicity is replaced by the metallicity difference ($\Delta$(O/H) with respect to the metallicity of galaxies with the same stellar mass but located on the Main Sequence. More specifically, in Figure \ref{figure:master plot}, $\Delta$(O/H) is plotted against total stellar mass and different curves on these plots are colour-coded with respect to the $\rm \Delta SFR_{MS}$, where we define $\rm \Delta SFR_{MS} = log(SFR) - log(SFR)_{MS}$, i.e. the difference between the SFRs with respect to the SFR of galaxies on the Main Sequence (MS) at a given stellar mass. The redder curves correspond to the galaxy populations with SFRs below the Main Sequence, while the bluer curves correspond to those galaxy populations whose SFRs are above the Main Sequence. 
 
 
 We find that the gas-phase metallicity of galaxy population with SFRs higher than MS galaxies are metal-poor compared to that of Main Sequence galaxies, and the metal deficiency becomes more prominent the further galaxies deviate from the Main Sequence, although the effect is less prominent at high stellar masses.
 On the other hand, the gas-phase metallicity of galaxies with SFR lower than the Main Sequence are more metal-rich, but primarily this effect is seen at low stellar masses; the metallicity excess increases with deviation from the Main Sequence, but steeply decreases with  increasing  stellar mass, becoming consistent with Main Sequence galaxies and then the difference increases again at high stellar masses for galaxies well below the Main Sequence.
 
 Note also that we are looking at the relative trends here where the relative uncertainty (due to binning) is much smaller than the absolute uncertainty on metallicity calibration, SFR and M$_{\star}$. Therefore, relative variation although small are quite significant.
 

 \section{Discussion and conclusion}
 \label{section:discussion}
 
 On the Main Sequence, galaxies are commonly believed to be in a
 quasi-steady equilibrium
 between inflows, outflows and star formation. An increase of star
 formation relative to the Main Sequence
 must be associated with
 either an increase of the additional gas from extra inflow, or an
 increased star formation efficiency.
 A decrease of star formation relative to the Main Sequence must be
 associated with either a shortage
 of gas due to removal by extra-outflows or due to processes \citep[e.g. halo heating,][]{Birnboim2007, vandevoort2011, Chruchill2012, Johansson2012} preventing gas accretion
 hence resulting in `starvation', or a decrease in star
 formation efficiency.
 
 \indent Recent works investigating the gas content in galaxies have revealed
 that deviations from the Main Sequence
 result from the variation in the gas content as well as in the
 star formation efficiency \citep[][]{Schreiber2016, Saintonge2017, Tacconi2018, Piotrowska2020, Zhang2019,Dou2020}. The metallicity of galaxies provides an additional dimension to
 investigate these aspects.
 For instance, as discussed in \citet{Mannucci2010} and in \citet{Bothwell2013, Bothwell2016a} the lower metallicity observed
 for highly star forming galaxies is likely associated with an
 increase of gas inflow that both boosts star formation
 and dilutes the gas-phase metallicity (an alternative explanation is
 that the increased star formation drives more powerful
 metal-loaded outflows, but the stronger anti-correlation with gas
 content favours the inflow-dilution scenario).
 As already mentioned in the introduction, \citet{Peng2015} and
 \citet{Trussler2020} show that the average
 {\it stellar} metallicity of passive and green-valley galaxies is higher
 than Main Sequence galaxies (especially for what concerns high redshift
 progenitors of passive galaxies) and that this can
 only be explained in terms of quenching being primarily associated with `starvation', i.e.
 suppression of gas inflow that on the one hand results into
 suppression of star formation (as the residual gas is being
 consumed by star formation), while on the other hand it results into an increase
 of the metallicity as the dilution effect of inflow
 is reduced.

 \indent We now have the opportunity to investigate the same effects by using
 the {\it gas} metallicity for galaxies spanning a broad
 range of offset from the Main Sequence, including galaxies not
 classified as star-forming in the BPT diagrams, hence enabling
 us to delve better into the green valley and quiescent region, where
 galaxies typically have LINER-like emission lines (which
 were generally excluded by previous studies of the FMR).

 \indent We can now interpret the metallicity trends shown in Fig \ref{figure:master plot} relative to Main Sequence galaxies in the context of the scenarios discussed above. Those galaxies with positive $\rm\Delta SFR_{MS}$, i.e. above the Main Sequence have metallicities lower than their counterparts on the Main Sequence. i.e. $\Delta$(O/H) is always $<0$. As mentioned above, this effect can be interpreted in terms of gas inflows dominating the galaxies lying above the Main Sequence, which both dilute the gas-phase metallicity and boost the star formation.
 
 The behaviour of galaxies below the
 Main Sequence (negative $\rm\Delta SFR_{MS}$) is more complex. At
 low masses the gas metallicity is clearly higher
 than in Main Sequence galaxies, therefore favouring a chemical evolution of these
 systems in which the metallicity increases as a consequence
 of lack of dilution from the inflow of fresh gas, i.e. starvation
 scenario. In these galaxies the chemical enrichment tend to proceed
 more in a closed-box or leaky-box scenario, in which galaxies are out
 of equilibrium and the metallicity increases more steeply
 than for the Main Sequence galaxies in which star formation and outflows
 are 
 balanced by inflows.
 Qualitatively, this is the same result obtained by \citet{Peng2015}
 and \citet{Trussler2020} for the stellar metallicities. However,
 quantitatively the result is quite different. The stellar metallicity
 difference between passive and Main Sequence galaxies (and between green valley
 and Main Sequence galaxies), as inferred from \cite{Trussler2020}, is illustrated in Fig. \ref{figure:master plot} with a thick red line (and with a thick green line for green valley galaxies). At low stellar masses the gas metallicity differences are similar to the stellar metallicity difference, supporting the same starvation scenario.
 However at high/intermediate stellar masses ($\gtrsim$ 10$^{10}$ M$_{\odot}$), although there is still some visible effect of the gas metallicity difference increasing with the deviation from the Main Sequence, quantitatively the gas metallicity differences are much smaller than the stellar metallicity differences.
 This is clearly indicating that the stellar population of massive
 galaxies below the Main Sequence has not formed out of
 gas with the same chemical properties as the gas that we are currently
 observing in these systems.
 
 A possible scenario is that in the case of the massive galaxies below the
 Main Sequence for which we can measure the gas-phase
 metallicity we are observing them in a peculiar evolutionary stage.
 Indeed, in order to measure the metallicity we need
 to detect nebular emission lines that, for passive/green-valley galaxies,
 region are typically characterized by LI(N)ER-like ratios,
 especially in the central region probed by the SDSS fibre \citep{Belfiore2017}. However, the majority of passive galaxies 
 are line-less, so for the majority of quiescent galaxies we cannot
 really measure the gas metallicity. Therefore, a likely scenario is that the
 small fraction of quiescent galaxies that have
 nebular emission (and for which we can measure the gas metallicity) may have an excess of gas because they have recently accreted gas from the CGM/IGM (hence typically with metallicity lower than the stellar metallicity
 of the accreting massive galaxy) and, therefore, in this case the observed gas is
 anomalously metal poor relative to their quiescent parent
 population. In a few cases this gas accretion may result
 into a rejuvenation event, with the accreted gas
 forming stars, as indeed observed in some massive galaxies \citep{Chauke2019}. However,
 as already emphasised, in most cases
 the accreted gas has LIER-like emission, i.e. not associated with star
 formation: it is likely ionized by the hard radiation
 field of the evolved (post-AGB) stars. Such hard radiation field is
 likely responsible for keeping the accreted gas
 hot and ionized and preventing it from forming stars. Such accreted gas may
 also likely be expelled at later times
 by the SNIa cumulative energy \citep[see][]{Trussler2020, Pipino2004,Matteucci2006} 
 and also by the energy
 injected by a weak AGN \citep{Ciotti2007, Ciotti2009}. 
 This effect of gas accretion with little SF associated with it, may
 also be responsible for the offset between stellar and
 gas metallicity for galaxies below the Main Sequence at low/intermediate masses. We also find that $\sim$78$\%$ of galaxies lying in the mass range 8 $<$ log(M$_{\star}$/M$_{\odot}$) $<$ 9, have $\Delta$(O/H) $>$ 0., implying that a majortiy of dwarf galaxies are going through a starvation phase rather than dilution via gas accretion.

 \indent Summarizing, our results support the scenario in which for galaxies
 above the Main Sequence the lower metallicity is associated with
 the dilution effect of gas accretion. For galaxies below the Main Sequence,
 at low masses their higher metallicity support the
 scenario in which their quenching is involving starvation (suppression
 of gas inflows);  outflows may also play a role
 in quenching, but starvation is necessary to account for the increased
 gas-phase metallicity relative to Main Sequence galaxies. At high masses
 the fact that galaxies below the Main Sequence have gas-phase metallicities
 closer to the Main Sequence galaxies, while
 their stellar metallicities are much higher than galaxies on the Main Sequence,
 suggests that we are observing these galaxies in a peculiar phase, in
 which they have recently accreted gas, but that such accreted
 gas is prevented from forming stars (in agreement with the independent
 findings that the efficiency of star formation decreases drastically
 toward the quiescent region). 
 
 \indent This study has allowed us to extend the FMR to the BP-non-SF galaxies in the local Universe, and is a step-forward
 in piecing together different evolutionary stages of BPT-non-SF galaxies
 dominated by different physical processes. Recent surveys have pointed out the presence of a substantial population of non-SF/quiescent galaxies in the high redshift Universe, though spectroscopy information on these populations
 is still sparse  
 \citep{McCracken2012,Straatman2016,Muzzin2013,Schreiber2018,  Morishita2019, Whitaker2013,Glazebrook2017, Santini2019, Merlin2018}.  The upcoming
 Multi-Object Optical and Near-infrared Spectrograph \citep{
 	Cirasuolo2020} on the VLT will enable the extensive spectroscopic
 follow-up of such imaging surveys \citep{Maiolino2020}, hence enabling us to explore the
 FMR for BPT-non-SF galaxies in the distant Universe, and will allow us to
 understand the physical processes driving quenching and chemical
 enrichment at different epochs in the history of the Universe.

\begin{acknowledgements}
 \indent  NK acknowledges the fellowship from the Schlumberger Foundation which supported her stay at KICC where most of the work was accomplished.  MC, RM and JT acknowledge ERC Advanced Grant 695671 ``QUENCH" and support by the Science and Technology Facilities Council (STFC). This research made use of the  on SDSS archival data, that can be retrieved from https://wwwmpa.mpa-garching.mpg.de/SDSS/DR7/ and Astropy, a community-developed core Python package for Astronomy \citep{Astropy2013}.

\end{acknowledgements}

%
\bibliographystyle{aa} 
\bibliography{biblio} 
%

\begin{appendix} 

\section{Alternative FMR using [SII]-BPT with and without AGNs}
\label{appendix:noagn}
Figure \ref{figure:no agn} shows the alternative representation of the FMR where star-forming and non-star-forming samples have been identified using the [S~\textsc{ii}]-BPT classification. The metallicities of the BPT-SF sample have been determined using O3N2 calibration from \citet{Curti2017} while for the BPT-non-SF sample, we applied the correction to these metallicities given by \citet{Kumari2019} for [S~\textsc{ii}]-BPT classification. The sample in the upper-panel also consists of AGNs while that in the lower-panel excludes AGNs. The overall behaviour of the FMR for both samples (with and without AGNs) selected on the basis of [S \textsc{ii}]-BPT is similar to that found in Figure \ref{figure:master plot} where [N \textsc{ii}]-BPT was used for sample selection. This shows that the FMR behaviour is independent of the inclusion or removal of AGNs. This strengthens our results from [N \textsc{ii}]-BPT classification where AGNs could not be excluded for a lack of recipe. We find however that metallicity differences for the massive galaxies in the [S\textsc{ii}]-BPT selected sample hardly go beyond 0 (Figure \ref{figure:no agn}) in contrast to the [N\textsc{ii}]-BPT selected sample (Figure \ref{figure:master plot}). This is likely because the initial O3N2 calibration from \citep{Curti2017} is devised from the [N \textsc{ii}]-BPT selected sample and not [S \textsc{ii}]-BPT selected sample.

\renewcommand{\thefigure}{A\arabic{figure}}
\begin{figure}
	\centering 
	\includegraphics[width=0.45\textwidth]{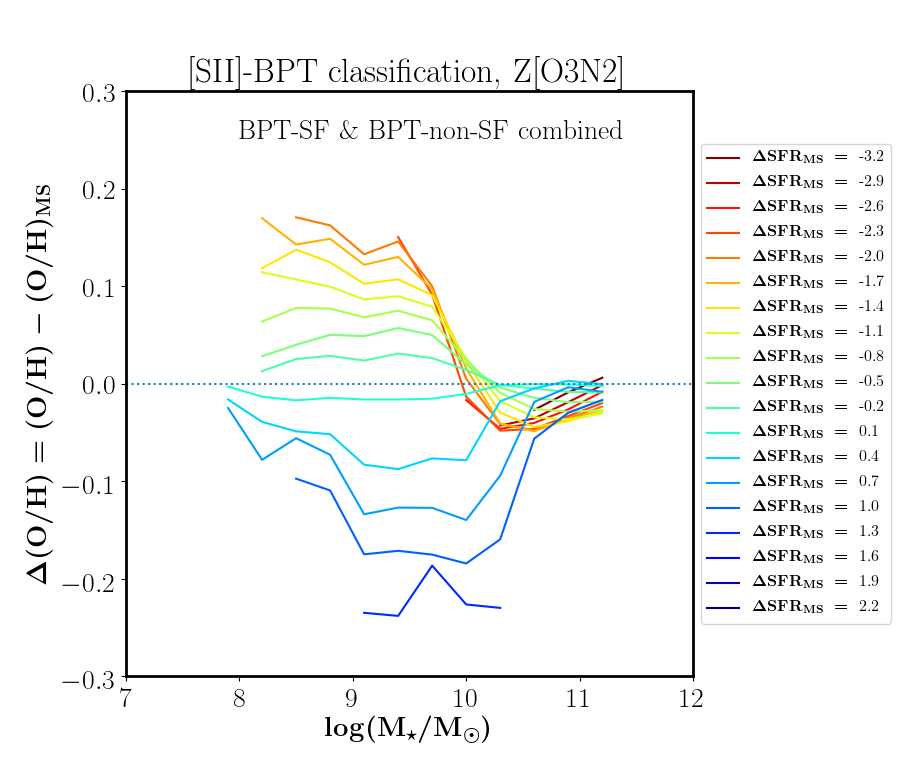}
	\includegraphics[width=0.45\textwidth]{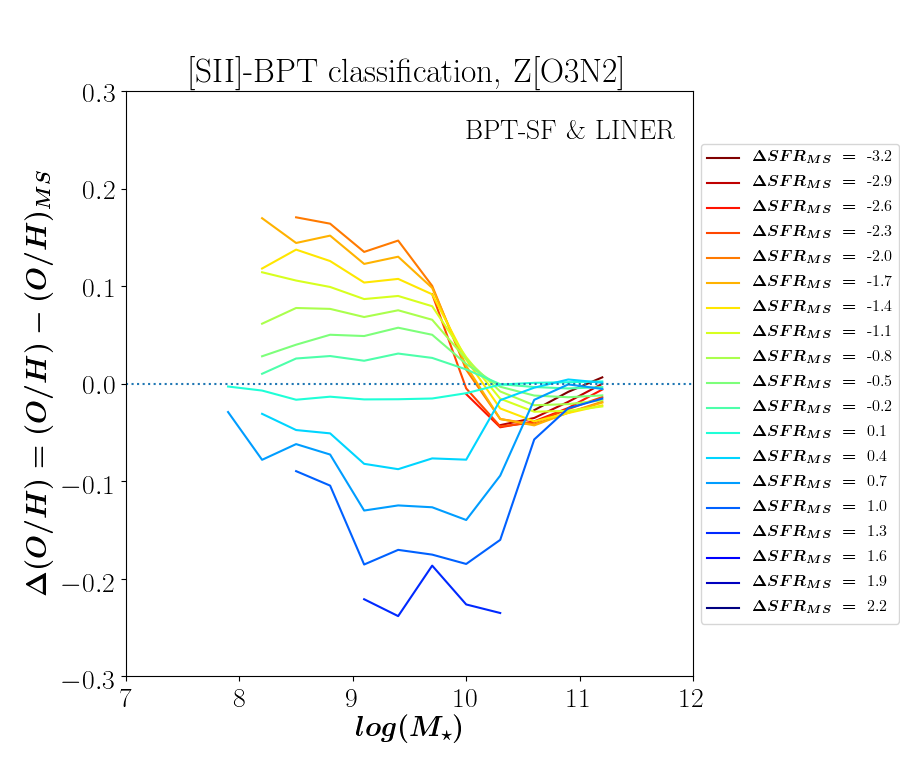}
	\caption{$\Delta$(O/H) versus stellar mass (log M$_{\star}$/M$_{\odot}$) for the entire sample comprising of BPT-SF and BPT-non-SF galaxies where [S~\textsc{ii}]-BPT is used for galaxy classification. The sample in the upper-panel includes AGNs while the sample in the lower panel excludes AGNs. The coloured curves on each plot correspond to median $\Delta$(O/H) estimated in bins of log M$_{\star}$/M$_{\odot}$ and $\rm \Delta SFR_{MS}$ and colour-coded with respect to $\rm \Delta SFR_{MS}$ in a given bin.}
	\label{figure:no agn}
\end{figure}

\end{appendix}

\end{document}